\def\Tr{{\text{Tr}}\,}

\def\be{\begin{equation}}
\def\ee{\end{equation}}
\def\bea{\begin{eqnarray}}
\def\eea{\end{eqnarray}}
\def\bse{\begin{subequations}}
\def\ese{\end{subequations}}

\documentclass[prl,twocolumn,showpacs,preprintnumbers,amsmath,amssymb]{revtex4}
\usepackage{graphicx}% Include figure files
\usepackage{dcolumn}% Align table columns on decimal point
\usepackage{bm}% bold math
\begin{document}
\preprint{}
\title{Nature of the quantum phase transition to a spin-nematic phase}
\author{T.R. Kirkpatrick$^{1}$ and D. Belitz$^{2}$}
\affiliation{$^{1}$Institute for Physical Science and Technology and Department
                   of Physics, University of Maryland, College Park, MD 20742\\
         $^{2}$Department of Physics and Theoretical Science Institute, University
                of Oregon, Eugene, OR 97403}
\date{\today}
\begin{abstract}
It is shown that the quantum phase transition in metallic non-s-wave
ferromagnets, or spin nematics, is generically of first order. This is due to a
coupling of the order parameter to soft electronic modes that play a role
analogous to that of the electromagnetic vector potential in a superconductor,
which leads to a fluctuation-induced first-order transition. A generalized
mean-field theory for the p-wave case is constructed that explicitly shows this
effect. Tricritical wings are predicted to appear in the phase diagram in a
spatially varying magnetic field, but not in a homogeneous one.
%
% 596 characters
\end{abstract}


\maketitle

Analogies between certain liquid-crystal phases (smectic, cholesteric, nematic,
columnar, blue) \cite{DeGennes_Prost_1993} and ordered states of electrons in
solids provide a suprising connection between soft condensed matter and
electronic systems. For instance, stripe phases in high-temperature
superconductors are analogous to smectics \cite{Kivelson_Fradkin_Emery_1998,
Kivelson_et_al_2003}, helical magnets are analogous to cholesterics
\cite{Belitz_Kirkpatrick_Rosch_2006a}, and electronic analogs of nematics can
be realized via Pomeranchuk instabilities of the Fermi surface
\cite{Oganesyan_Kivelson_Fradkin_2001}. Stripe and nematic order can occur in
the spin channel as well as in the charge channel, and spin nematics provide
magnetic analogs of non-s-wave superconductors that have been invoked to
explain the enigmatic behavior of Sr$_3$Ru$_2$O$_7$ \cite{Kee_Kim_2005,
Ho_Schofield_2008} and the `hidden order' in URu$_2$Si$_2$
\cite{Varma_Zhu_2006}.
% At a more technical level, analogies emerge that are
% even more remarkable. For instance, it has long been known that the transition
% in liquid crystals from the nematic phase to the smectic-A phase is described
% by a theory that is very similar to the one describing the transition from a
% normal metal to a BCS superconductors, and that the nematic Goldstone modes in
% the former play a role in determining the nature of the transition that is
% analogous to that of the electromagnetic vector potential in the latter
% \cite{Halperin_Lubensky_Ma_1974}.

For a theoretical description of phase transitions, Landau theory is a standard
tool \cite{Landau_Lifshitz_V_1980}. Once an order parameter (OP) $\phi$ has
been identified one constructs all terms in the free energy that are consistent
with the general symmetry properties of the system. Replacing the OP field by a
constant yields the most general mean-field theory for the phase transition.
Fluctuations of the OP can also be considered, which leads to the
Landau-Ginzburg-Wilson (LGW) framework that is amenable to an analysis by
renormalization-group methods \cite{Wilson_Kogut_1974}. Landau theory is
applicable to both classical transitions that are driven by thermal
fluctuations and tuned by changing the temperature $T$, and quantum transitions
at zero temperature that are driven by quantum fluctuations and tuned by a
non-thermal parameter such as external pressure or composition
\cite{Sachdev_1999}.

\begin{figure}[t]
\vskip -0mm
\includegraphics[width=5.0cm]{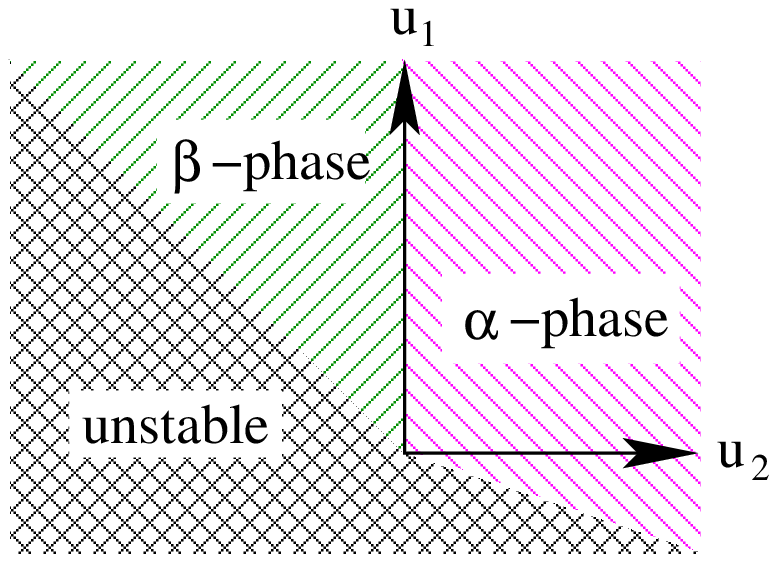}
\caption{Phase diagram for the Landau free energy given in Eq.\ (\ref{eq:1}).
 The $\alpha$ and $\beta$-phases are stable for $u_2>0$, $3u_1>-u_2$, and
 $u_2<0$, $u_1>-u_2$, respectively. The Landau theory predicts a second-order
 transition into either phase at $t=0$.
 %In the hatched region the system is unstable.
 }
\label{fig:1}
\end{figure}

In order to experimentally check the proposals concerning liquid-crystal
analogs in electronic systems it is important to predict as many qualitative
features of this exotic order as possible. To this end a comprehensive analysis
of the Landau theory for an electronic nematic state has been developed, and
the quantum phase transition has been analyzed in analogy to Hertz's theory of
s-wave quantum ferromagnets \cite{Hertz_1976}, see Ref.\
\onlinecite{Wu_et_al_2007} and references therein. The theory has been
developed for both the spin and the charge channel; here we will focus on the
spin channel and especially on the p-wave case. The OP for a non-s-wave
ferromagnet is of the form
$\langle{\bar\psi}_a(x)\,\sigma^i_{ab}f(\partial_{\bm x})\,\psi_b(x)\rangle$,
with $\sigma^{1,2,3}$ the Pauli matrices. $\psi =
(\psi_{\uparrow},\psi_{\downarrow})$ is a fermionic spinor field, $\bar\psi$ is
its adjoint, $x = ({\bm x},\tau)$ comprises the real space position ${\bm x}$
and the imaginary time variable $\tau$, $f$ is a tensor-valued monomial
function of the gradient operator, and $\langle \ldots \rangle$ denotes the
quantum mechanical and thermodynamic average. In the p-wave case the function
$f$ is linear in the gradient and the OP field thus carries a spin index $i$
and an orbital index $\alpha$ ($i=1,2,3$, $\alpha = 1,2,3$). We will denote it
by $N_{i,\alpha}(x)$ which in Landau becomes a number $N_{i,\alpha}$. The most
general Landau free energy up to quartic order in $N_{i,\alpha}$ has the form
\cite{Wu_et_al_2007}
\be
F_{\text{L}} = t\,N_{i}^{\alpha} N^{i}_{\alpha} + u_1\,\left(N_{i}^{\alpha}
N^{i}_{\alpha}\right)^2 + u_2\,N_{i}^{\alpha} N^{i}_{\beta}\,N^{j}_{\alpha}
N_{j}^{\beta},
\label{eq:1}
\ee
with a summation convention implied. Depending on the relative values of the
Landau parameters $u_1$ and $u_2$ there are two distinct phases, the
$\alpha$-phase and the $\beta$-phase \cite{Wu_et_al_2007}. They are
characterized by $N_{i,\alpha} = n_{\alpha} {\hat N}_i$ for the $\alpha$-phase,
with ${\hat N}$ a unit vector in spin space, and $N_{i,\alpha} = N\,{\hat
e}^{(\alpha)}_i$ for the $\beta$-phase, with $N$ a number and ${\hat
e}^{(1,2,3)}$ three mutually orthogonal unit vectors. The Landau theory
predicts a second-order transition into either phase at $t=0$. The
corresponding phase diagram is shown in Fig.\ \ref{fig:1}. For the classical
transition at $T>0$ in 3-d fluctuations of the OP will modify the mean-field
critical behavior predicted by Eq.\ (\ref{eq:1}). For the corresponding quantum
phase transition the dynamical critical exponent stabilizes the Gaussian fixed
point in 3-d and the LGW theory that generalizes Eq.\ (\ref{eq:1}) predicts the
transition to be second order with mean-field critical behavior
\cite{Wu_et_al_2007}.

There are reasons to doubt the validity of the latter prediction. Despite its
generality, Landau theory hinges on certain implicit assumptions. One is that
an expansion of the free energy in powers of the OP is well behaved, i.e., that
the coefficients in a Taylor expansion are finite. The analogous statement
within an LGW framework is that the field theory is local. This assumption can
be violated if there are soft modes other than the OP fluctuations that couple
to the OP. In such a situation, the construction of a Landau theory entirely in
terms of the OP amounts to integrating out these soft modes, which may lead to
a non-local theory. Technically, a nonzero OP can give the additional soft
mode(s) a mass, which implies that the free energy as a function of the OP
cannot be analytic at $\phi = 0$. The resulting changes to the phase transition
depend on the properties of the original Landau theory and on the sign of the
leading nonanalytic term. A classic example is the case of a type-I or weakly
type-II BCS superconductor, where a simple Landau theory predicts a
second-order or continuous transition. However, the coupling of the OP to the
electromagnetic vector potential leads to an effective free energy that
contains a term of cubic order in the OP, which leads to a first-order phase
transition \cite{Halperin_Lubensky_Ma_1974}. The same is true for the
nematic-to-smectic-A transition in liquid crystals, with the nematic Goldstone
modes playing a role analogous to that of the vector potential. This phenomenon
is known as a fluctuation-induced first-order transition in condensed matter
physics, and as the Coleman-Weinberg mechanism for mass generation in particle
physics \cite{Coleman_Weinberg_1973}. Another possibility is that the
transition remains second order, but the coupling to the additional soft modes
changes the universality class. This is believed to be the case deep in the
type-II region \cite{Dasgupta_Halperin_1981, Herbut_Yethiraj_Bechhoefer_2001}.

For the quantum phase transition from a Fermi liquid to an electronic nematic
state there are soft modes in addition to the OP fluctuations, namely,
particle-hole excitations that are soft at $T=0$ and acquire a mass at nonzero
$T$. The question is whether these soft modes couple to the OP in a way that
invalidates simple Landau theory. In this Letter we show that they do for spin
nematics (but not for charge nematics) and generically lead to a quantum phase
transition that is fluctuation-induced first order. For simplicity, we will
discuss the p-wave case in three dimensions (3-d) unless noted otherwise.

We will first state and discuss our results, and then sketch their derivation.
In a generalized mean-field theory analogous to Ref.\
\onlinecite{Halperin_Lubensky_Ma_1974} that neglects order-parameter
fluctuations, but keeps the fermionic degrees of freedom and their coupling to
the OP in a Gaussian approximation one finds, for a $3$-$d$ system, an equation
of state whose qualitative features are represented by
\be
h_N = t\,N + v\,N^3 \ln(N^2 + T^2) + u\,N^3 + o(N^3).
\label{eq:2}
\ee
Here $N$ is the number-valued OP for the $\alpha$-phase (the result for the
$\beta$-phase is structurally the same), $u = u_1 + u_2$, and $h_N$ is the
field conjugate to $N$ (more on this below). $v>0$ is a positive definite
Landau coefficient that is given in terms of spin-triplet interaction
amplitudes. $o(x)$ denotes terms smaller than $x$ as $x \to 0$. The Landau
parameters $t$ and $u$ are also (weakly) $T$-dependent, but we explicitly show
only the $T$-dependence of the logarithmic term since it cuts off a nonanalytic
dependence on $N$. For $1<d<3$ the corresponding nonanalyticity at $T=0$ is
given by $N^d$ instead of $N^3\ln N$.

\begin{figure}[t]
\vskip -0mm
\includegraphics[width=7.0cm]{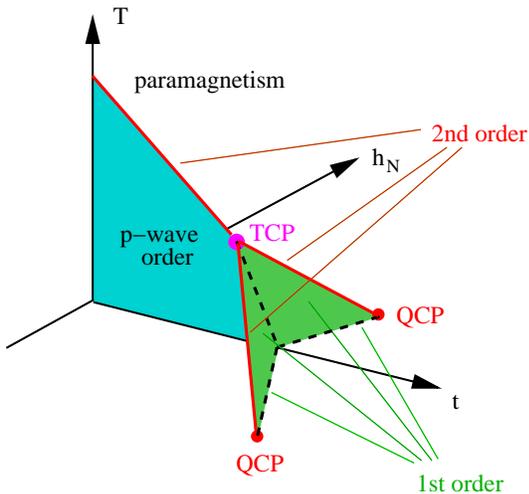}
\caption{Schematic phase diagram in the space spanned by $T$, $h_N$,
 and the non-thermal control parameter $t$. Shown are the region of p-wave order,
 the tricritical point (TCP), the two quantum critical
 points (QCP), and the various phase transition lines and surfaces. For $h_N = 0$
 the transition is first order above the TCP and second order below. For
 $h_N \neq 0$ tricritical wings connect the TCP and
 the two QCPs.
 % $h_N$ can be realized by means of a non-homogeneous magnetic
 % field as explained in the text.
}
\label{fig:2}
\end{figure}

Equation (\ref{eq:2}) predicts several features that are qualitatively
different from the Landau theory of Ref.\ \onlinecite{Wu_et_al_2007}. At $T=0$,
the term $N^3\ln N$ is negative and larger than $N^3$, which drives the
transition first order below a tricritical temperature $T_{\rm{tc}} =
\exp{(-u/2v)}$. At $T=0$ there is a first-order transition at $t = t_1 > 0$
that preempts the second-order transition predicted by Landau theory and leads
to a discontinuous change of $N$ from $0$ to $N_1 = \exp[-(1+u/v)/2]$. For
$t_1$ one finds $t_1 = v\,N_1^2/4$. In the presence of a conjugate field,
tricritical wings appear in the parameter space spanned by $T$, $h_N$, and $t$,
as is the case at any tricritical point \cite{Griffiths_1970}. As a result, we
predict the phase diagram to have the same qualitative structure as the one
observed for quantum ferromagnets \cite{Pfleiderer_Julian_Lonzarich_2001,
Tafour_et_al_2010}, see Fig.\ \ref{fig:2}.
There are surfaces (``wings'') of first-order transitions that are bounded by
second-order transitions and end in a pair of quantum critical points in the
$T=0$ plane. The quantum critical exponents at these points are mean-field
like. We choose as independent the static exponents $\beta$ and $\delta$, and
the dynamical exponent $z$, i.e., the scale dimension $[T]_{\text{c}}$ of $T$
at criticality. We find
\be
\beta = 1/2\quad,\quad\delta = 3\quad,\quad z=3\ .
\label{eq:3}
\ee
These exponents are {\em exact}, since the conjugate field renders massive the
fermionic soft modes that, in the absence of a field, lead to the first-order
transition described above. $\beta$ and $\delta$ govern the dependence of $N$
on $t$ and $h_N$, respectively and $z$ determines the temperature scaling of,
e.g., the specific heat. The $T$-dependence of the OP at criticality, however,
is {\em not} given by $z$ for the same reason as in the s-wave case
\cite{Millis_1993, Sachdev_1997}. It is determined by a fluctuation scale
$[T]_{\text{fluc}} = 9/(d+1)$ that yields the leading $T$-dependence for all
$d<5$. For $d = 3$ we find for the deviation $\delta N$ of $N$ from its value
at the critical point
\be
\delta N (t_{\text c}, h_{N}^{\text c}, T) \propto - T^{4/9},
\label{eq:4}
\ee
which also is exact. The continuous transitions at $T>0$, on the other hand,
will be modified by OP fluctuations.
%; e.g., the continuous transition
%from a paramagnet to a p-wave ferromagnet above the tricritical point is in the
%universality class of a classical transition with $O(3)\times SU(2)$ symmetry.

In $d=2$ there is no long-range order at $T>0$. However, at $T=0$ one still has
a first-order quantum phase transition, with $N_1 = (3v/4u)^2$, $t_1 =
u\,N_1^2/6$.
%\be
%N_1 = (3v/4u)^2\quad,\quad t_1 = u\,N_1^2/6.
%\label{eq:5}
%\ee

These predictions are all analogous to the properties of quantum s-wave
ferromagnets \cite{Belitz_Kirkpatrick_Vojta_1999,
Pfleiderer_Julian_Lonzarich_2001, Belitz_Kirkpatrick_Rollbuehler_2005,
Tafour_et_al_2010}. A crucial difference that allows to distinguish the two
cases is the nature of the conjugate field. In the s-wave case, it is the
physical magnetic field. In the p-wave case (as well as for all higher angular
momenta) a homogeneous magnetic field does not couple to the OP.
% and thus has no effect.
However, a conjugate field $h_N$ is induced by a non-homogeneous magnetic
field. For instance, a magnetic field ${\bm h}({\bm x})$ with components $h_i$
in two independent directions induces a conjugate field
\be
h_N^{i,\alpha}({\bm x}) \propto \epsilon_{ijk}\ \frac{1}{V} \int d{\bm y}\
h_j({\bm y})\,\partial_{\alpha} h_k({\bm x})
\label{eq:6}
\ee
with $V$ the system volume and a prefactor proportional to the p-wave
interaction amplitude. A homogeneous field $h_N$ results if $h_k({\bm x})$ is a
linear function of position.

To summarize, our theory of low-$T$ p-wave ferromagnets or magnetic nematics
predicts a first-order transition at asymptotically low $T$ that is separated
from a second-order transition at higher $T$ by a tricritical point. In this
respect it behaves the same way as an ordinary s-wave ferromagnet. However, in
contrast to the s-wave case a p-wave ferromagnet is unaffected by a homogeneous
magnetic field, whereas in a properly designed inhomogeneous field it shows the
same striking tricritical-wing structure found in s-wave ferromagnets
\cite{Pfleiderer_Julian_Lonzarich_2001, Belitz_Kirkpatrick_Rollbuehler_2005,
Tafour_et_al_2010}.

A derivation of these results hinges on an effective theory of soft modes in
clean electron systems that will be described in detail elsewhere
\cite{us_tpb}; here we just sketch some salient features. The building blocks
are bilinear products ${\bar\psi}_{n_1}({\bm x})\,\psi_{n_2}({\bm y})$ that can
be constrained to bosonic, quaternion-valued, fields $Q_{n_1 n_2}({\bm x},{\bm
y})$. Here $n_1$ and $n_2$ label fermionic Matsubara frequencies $\omega_n =
2\pi T(n+1/2)$. This was pioneered by Wegner in the context of noninteracting
electronic systems with quenched disorder \cite{Wegner_1979,
Wegner_Schaefer_1980}, generalized by Finkel'stein to interacting electrons
\cite{Finkelstein_1983, Finkelstein_1984b}, and later elaborated on by the
present authors \cite{Belitz_Kirkpatrick_1997}. These theories were all
formulated in terms of local density matrix fields $Q_{n_1 n_2}({\bm x},{\bm
x})$. Here we need to generalize to non-local phase-space variables $Q_{n_1
n_2}({\bm x},{\bm y})$. For a spherical Fermi surface we expand the Fourier
transform of the $Q$ in spherical harmonics $Y_l^m$,
\be
Q_{n_1 n_2}^{l,m}({\bm k}) = \frac{\sqrt{4\pi}/V}{\sqrt{2l+1}} \sum_{\bm p}
Y_l^m(\Omega_{\bm p})\, Q_{n_1 n_2}({\bm p}+{\bm k}/2,{\bm p}-{\bm k}/2),
\label{eq:7}
\ee
with $\Omega_{\bm p}$ the solid angle for the wave vector ${\bm p}$. For our
purposes we need an effective theory for soft modes in clean systems. This is a
harder problem than the disordered one since in clean systems there are many
more soft modes, namely, {\em all} moments of the phase-space excitations, not
just the zeroth one. A Ward identity shows that all $Q_{n_1 n_2}^{l,m}({\bm
k})$ with $n_1\,n_2<0$ are soft. They are the Goldstone modes of a symmetry
between retarded and advanced degrees of freedom that is spontaneously broken
if the Fermi energy lies within a band; we denote them by $q_{n_1
n_2}^{l,m}({\bm k})$ \cite{soft_modes_footnote}. The role of the $q$ in the
present theory is analogous to that of the vector potential in the case of the
BCS transition in Ref.\ \onlinecite{Halperin_Lubensky_Ma_1974}. The various
angular momentum channels, $l=0,1,\ldots$, couple and all $q$-correlation
functions scale as the inverse of the wave number which in turn scales as a
frequency. In a schematic notation, leaving out all constant prefactors,
\be
\langle q_{n_1 n_2}^{l_1,m_1}({\bm k}) q_{n_3 n_4}^{l_2,-m_1}(-{\bm k}) \rangle
\propto \frac{\delta_{n_1 - n_2,n_3 - n_4}}{\vert{\bm k}\vert + \Omega_{n_1 -
n_2}}
\label{eq:8}
\ee
with $\Omega_n = 2\pi Tn$ a bosonic Matsubara frequency. This structure holds
for both noninteracting and interacting electron systems, which reflects the
fact that the system in the absence of magnetic order is a Fermi liquid. By
integrating out the massive degrees of freedom in a conserving approximation
one can construct an effective soft-mode theory in terms of the $q$ only that
is analogous to, albeit more complex than, the generalized nonlinear
sigma-model for disordered interacting electrons \cite{Finkelstein_1983,
Belitz_Kirkpatrick_1997}. The generalized mean-field theory discussed above can
be derived analogously to the corresponding theory for the s-wave case
\cite{Belitz_Kirkpatrick_Vojta_1999, Kirkpatrick_Belitz_2003b}. The leading
coupling between the static, homogeneous OP $N$ and the fermionic soft modes
$q$ takes the form $N\,\Tr[q({\bm k})\,q^{\dagger}({\bm k})]$ where $\Tr$
traces over the angular momentum, frequency, and spin degrees of freedom. To
leading order in $N$ and $q$ one finds a fermionic part of the action with the
schematic structure
\be
{\cal A}_{q} \propto \Tr [(\vert{\bm k}\vert + \Omega + N)\,q({\bm
k})\,q^{\dagger}({\bm k})].
\label{eq:9}
\ee
Upon integrating out $q$ one obtains a contribution to the equation of state
that is schematically given by
\be
I(N,T) \propto \frac{1}{V}\sum_{\bm k} T\sum_{n=1}^{\infty}
\frac{N\,k^2\,\Omega_n^2} {(k^2+\Omega_n^2)^2 (k+\Omega_n)^2 + N^2 k^2
\Omega_n^2}
\label{eq:10}
\ee
with a positive prefactor. $I(N,T)$ appears in the equation of state in
addition to the standard Landau terms.

Asymptotic analysis reveals that the singularity at $N=0$, $T=0$ takes the form
\bse
\label{eqs:11}
\bea
&&\hskip -20pt I(N,T=0) \propto N \begin{cases} {\rm const.} - N^{(d-1)}\quad
(1<d<3), \cr
                       {\rm const.} + N^2 \ln N  \ \ (d=3),
         \end{cases}
\label{eq:11a}\\
&&\hskip -20pt \left. I(N,T)/N\right|_{N\to 0} \propto {\rm const.} +
T^{d-1}\quad (1<d\leq 3).
\label{eq:11b}
\eea
\ese
For $d=3$, and neglecting the analytic $T$-dependence of the term linear in
$N$, the $v$-term Eq.\ (\ref{eq:2}) adequately represents the coupling of the
soft fermionic modes to the OP (note the absence of a $T^2\ln T$ term in
$d=3$).

We close with four remarks. First, the coupling of $O(Nqq^{\dagger})$ that
generates the singular dependence on $N$ is missing from the Hertz-type theory
of Ref.\ \onlinecite{Wu_et_al_2007}. It amounts to taking into account
fermionic loops in deriving the LGW theory, whereas Hertz theory treats the
fermions in a tree approximation. Second, our results do {\em not} carry over
to charge nematics. In that case the OP does not give the $q$ a mass and the
Hertz-type LGW theory \cite{Wu_et_al_2007} is valid, consistent with a general
argument first given in Ref.\ \onlinecite{Belitz_Kirkpatrick_Vojta_2002}.
Third, the generalized mean-field theory still contains several approximations.
The Gaussian approximation for the $q$ is of no qualitative consequence;
Fermi-liquid theory ensures that the exact propagators have the same scaling
properties as Eq.\ (\ref{eq:8}). Higher-order coupling terms between $N$ and
the $q$ are irrelevant in a renormalization-group sense and keeping only Eq.\
(\ref{eq:9}) suffices. This leaves the mean-field approximation for the OP.
This becomes exact in a well-defined limit. $N\neq 0$ gives the $q$-propagator
a mass, see Eq.\ (\ref{eq:9}), which defines a length scale $\lambda \propto
1/N$. Together with the magnetic correlation length $\xi$ this defines a
Ginzburg parameter $\kappa = \lambda/\xi$. For $\kappa \to 0$ (extreme type-I
case) OP fluctuations are negligible and low-$T$ transition inevitably is first
order. For $\kappa > 0$, and especially for $\kappa >> 1$ (type-II limit), the
role of OP fluctuations must be examined. This is a difficult problem that has
been studied for classical transitions \cite{Dasgupta_Halperin_1981,
Herbut_Yethiraj_Bechhoefer_2001}, and to some extent for the quantum s-wave
ferromagnetic one \cite{Kirkpatrick_Belitz_2003b}, but more work is needed on
magnetic transitions in this limit. Fourth, all of our conclusions also hold
for spin nematics in higher angular momentum channels, only the realization of
the conjugate field changes.

This work was supported by the National Science Foundation under Grant Nos.
DMR-09-29966, and DMR-09-01907.

%\bibliography{p_wave}

\begin{thebibliography}{33}
\bibitem[{sof()}]{soft_modes_footnote}
\bibinfo{note}{In a Fermi liquid, these phase-space flucutations can be
  identified with particle-hole excitations. They will, however, remain soft in
  non-Fermi liquids as well.}
\end{thebibliography}

\end{document}